\documentclass[aps,prd,twocolumn,superscriptaddress,nofootinbib]{revtex4-2}

\usepackage{amsmath,amssymb,bm}
\usepackage{graphicx}
\usepackage{booktabs}
\usepackage{xcolor}
\usepackage{hyperref}
\usepackage{siunitx}
\usepackage{mathtools}
\usepackage{placeins}
\DeclareMathOperator{\arccot}{arccot}

\makeatletter
\newenvironment{fullwidthhere}{%
  \par\ignorespaces
  \onecolumngrid
  \vskip6\p@
}{%
  \par\vskip6\p@
  \twocolumngrid\global\@ignoretrue
  \@endpetrue
}
\makeatother

\hypersetup{colorlinks=true,citecolor=blue,linkcolor=blue,urlcolor=blue}
\newcommand{\md}{m_{\rm D}}
\newcommand{\nuhat}{\widehat\nu}
\newcommand{\Ecal}{\mathcal E}
\newcommand{\VinfR}{V_\infty^{R}}

\begin{document}

\title{Complex-energy spectrum and dissociation of heavy quarkonia in an anisotropic collisional quark-gluon plasma}

\author{Najmul Haque}
\email{nhaque@niser.ac.in}
\affiliation{School of Physical Sciences, National Institute of Science Education and Research, An OCC of Homi Bhabha National Institute, Jatni, Khurda 752050, India}

\begin{abstract}
We solve the Schr\"odinger equation for heavy quarkonia using the full complex potential of a momentum-anisotropic, collisional quark-gluon plasma.  The real and imaginary parts of the complex eigenvalue give the in-medium binding energy and thermal width, while the corresponding eigenfunction determines the spatial distribution of the state.  The noncentral anisotropic interaction is treated by expanding the potential in Legendre multipoles and solving the resulting coupled radial equations at fixed magnetic quantum number.  The collision rate is tied to the running coupling, with the pinch singularity of the imaginary potential used to restrict the allowed parameter domain.  Along the resulting pinch-free trajectories, the dissociation temperature of the $1S$ charmonium and bottomonium states increases with anisotropy.  The widths obtained from the full complex eigenvalue differ appreciably from first-order estimates based on the imaginary potential, while the corresponding $1S$ eigenfunctions remain dominated by their leading $S$-wave component.  We compare the thermal widths with available lattice-QCD results and use them to illustrate a static survival probability.
\end{abstract}

\maketitle

\section{Introduction}
\label{sec:intro}

Heavy quarkonia have long been used as probes of deconfined QCD matter.  The large heavy-quark mass makes the internal motion of a $Q\bar Q$ pair predominantly nonrelativistic and permits a potential description over a useful range of scales.  Matsui and Satz first proposed charmonium suppression as a consequence of color screening in a quark-gluon plasma (QGP)~\cite{Matsui:1986dk}.  Since then, quarkonium suppression and regeneration have become central observables in heavy-ion phenomenology; a broad overview is given in Ref.~\cite{Rothkopf:2019ipj}.

Potential models provide a direct connection between the in-medium heavy-quark interaction and the quarkonium spectrum.  Early applications focused on screening of the real part of the interaction and the associated reduction of the binding energy~\cite{Agotiya:2008ie}.  Real-time finite-temperature calculations later showed that the static potential is complex: its imaginary part arises from Landau damping and scattering processes and gives the state a thermal width~\cite{Laine:2006ns,Brambilla:2008cx}.  Complex potentials have since been used in direct Schr\"odinger calculations~\cite{Margotta2011,Thakur:2013nia} and in lattice-informed studies of in-medium spectral properties~\cite{Burnier2015}.  Open-quantum-system approaches provide a complementary description in which decoherence, dissociation, and recombination are treated dynamically~\cite{BrambillaOQS2017,Akamatsu:2020ypb}.

A useful point of comparison for the present calculation is the potential-model analysis of Srivastava, Chaturvedi, and Thakur~\cite{Srivastava:2018vxp}.  For an isotropic, collisionless plasma, they solved the Schr\"odinger equation with the real part of the potential, used the resulting temperature-dependent wave functions to evaluate the imaginary part perturbatively, and constructed survival probabilities from the corresponding widths.  Here we instead diagonalize the Hamiltonian containing $V_R+iV_I$ from the outset.  The binding energy, width, and medium-modified wave function therefore follow from the same complex eigenvalue problem.

Several extensions of the potential picture have addressed situations relevant to an evolving QGP.  Momentum-space anisotropy, expected during rapid longitudinal expansion~\cite{Romatschke:2003ms}, modifies both screening and damping; its consequences for quarkonium have been studied in Refs.~\cite{Thakur:2012eb,Dumitru:2009fy,Nopoush:2017zbu,Strickland:2011aa}.  Related studies have considered quarkonium motion through the medium~\cite{Thakur:2016cki,Sebastian:2022sga}, strong magnetic fields~\cite{Singh:2017nfa,Sebastian:2023tlw}, bulk-viscous corrections~\cite{Thakur:2020ifi,Thakur:2021vbo}, and modifications associated with the Gribov--Zwanziger action~\cite{Debnath:2023dhs}.  These examples emphasize that the spectrum and width can depend on more than an equilibrium Debye mass.

In Ref.~\cite{Debnath2026}, Debnath, Thakur, and I derived the complex heavy-quark potential in a spheroidally anisotropic plasma with a conserving Bhatnagar--Gross--Krook (BGK) collision kernel.  The same analysis identified a Weibel-instability-related double pole in the imaginary part of the potential and showed how a finite collision rate can move the system outside the singular region.  The widths in that work were estimated perturbatively using a prescribed Coulombic wave function.

The aim here is to solve the corresponding bound-state problem without treating the imaginary part as a perturbation.  Because the anisotropic interaction is axially symmetric but noncentral, the potential is expanded in Legendre multipoles and the Schr\"odinger equation is solved as a coupled partial-wave problem at fixed magnetic quantum number $m$.  This formulation retains the angular kinetic energy and partial-wave mixing while giving the complex pole and wave function in a single calculation.

\section{Medium inputs and the complex potential}
\label{sec:potential}

\subsection{Running coupling and Debye mass}

The hard scale of the spheroidal distribution is denoted by $\lambda$.  To close the running quantities in the present calculation we identify this scale with the temperature argument used in the perturbative input, $\lambda\equiv T$.  In the anisotropic system this should be understood as an operational hard-scale matching prescription rather than a statement of exact thermal equilibrium.

We use the one-loop running-coupling prescription employed in Refs.~\cite{Haque:2014rua,Haque:2024gva}, with the renormalization scale set to $\Lambda=2\pi T$,
\begin{equation}
\alpha_s(T)=\frac{12\pi}{(11N_c-2N_f)\ln[(2\pi T)^2/\Lambda_{\overline{\rm MS}}^2]},
\label{eq:alphas}
\end{equation}
and take $N_c=N_f=3$ and $\Lambda_{\overline{\rm MS}}=0.176$ GeV.  For the Debye screening mass we use the leading-order thermal expression
\begin{equation}
\md^2(T)=g^2(T)T^2\left(\frac{N_c}{3}+\frac{N_f}{6}\right).
\label{eq:md}
\end{equation}
where $g^2(T)=4\pi\alpha_s(T)$.

\subsection{Collision rate}
\label{sec:collision}

A leading-logarithmic estimate of the parton interaction rate gives~\cite{Thoma:1993vs}
\begin{equation}
\nu \approx 5.2\,\alpha_s^2 T\,
\ln\!\left(\frac{0.25}{\alpha_s}\right).
\label{eq:nuLL}
\end{equation}
At couplings relevant for heavy-ion collisions the logarithm in this expression can become negative, signaling the breakdown of the strict leading-log scale hierarchy.  Following Ref.~\cite{Schenke:2006xu}, we therefore retain an order-one constant inside the Coulomb logarithm.  Dividing Eq.~\eqref{eq:nuLL} by the Debye mass in Eq.~\eqref{eq:md} gives
\begin{equation}
C_\nu(N_c,N_f)=
\frac{5.2}{\sqrt{4\pi\left(N_c/3+N_f/6\right)}}.
\label{eq:Cnu}
\end{equation}
For the present $N_c=N_f=3$ calculation, $C_\nu=5.2/\sqrt{6\pi}=1.1977\simeq1.20$, and the collision rate is therefore
\begin{equation}
\frac{\nu(T;c)}{\md(T)}
=C_\nu\,\alpha_s(T)^{3/2}
\ln\!\left(c+\frac{0.25}{\alpha_s(T)}\right).
\label{eq:nu}
\end{equation}
We take $1\le c\le2$, with $c=1.5$ as the central value, whenever this interval is compatible with the pinch-free domain of the imaginary potential.  Let $\mathcal T$ denote the set of temperatures included in the medium scan; in the present calculation it spans $0.15\le T\le0.80$ GeV.  At finite anisotropy the lower edge is increased only if required by the pinch-free condition:
\begin{align}
c_{\rm low}(\xi)&=\max\!\left[1,\,\max_{T\in\mathcal T}c_{\min}(T,\xi)+\delta c\right],\nonumber\\
c_{\rm cen}(\xi)&=c_{\rm low}(\xi)+0.5,\qquad
c_{\rm high}(\xi)=c_{\rm low}(\xi)+1,
\label{eq:cscheduledef}
\end{align}
where $c_{\min}(T,\xi)$, defined in Eq.~\eqref{eq:cmin}, is the value required by the pinch boundary at a given $(T,\xi)$, and $\delta c=0.02$ is a small additive safety padding applied when the preferred band must be shifted upward.  No value of $c$ is fitted to the quarkonium spectrum.

\subsection{Anisotropic real and imaginary potentials}

We use the complex potential derived in Ref.~\cite{Debnath2026},
\begin{equation}
V(\bm r;T,\xi,c)=V_R(\bm r;T,\xi,c)+iV_I(\bm r;T,\xi,c),
\label{eq:complexV}
\end{equation}
where $\nu(T;c)$ is fixed by Eq.~\eqref{eq:nu}.  The anisotropy direction is chosen as the $z$ axis, $\mu=\cos\Theta=\hat{\bm r}\cdot\hat{\bm n}$, and $t=\cos\theta=\hat{\bm p}\cdot\hat{\bm n}$.  The Cornell input is
\begin{equation}
V_C(r)=-\frac{\alpha}{r}+\sigma r,
\qquad \alpha=\frac43\alpha_s(T),
\qquad \sigma=(0.44\,{\rm GeV})^2.
\end{equation}

\subsubsection{Static mass scales and fluctuation function}

For completeness, we list explicitly the static quantities entering the real and imaginary parts of the potential~\cite{Debnath2026}.  We define
\begin{subequations}
\begin{align}
f_0(\xi)&=\frac{1}{1+\xi}+\frac{\arctan\sqrt{\xi}}{\sqrt{\xi}},\\
f_1(\xi)&=-\frac{1}{\xi(1+\xi)}+\frac{\arctan\sqrt{\xi}}{\xi^{3/2}},
\end{align}
\end{subequations}
and introduce the static master integrals
\begin{equation}
\widetilde{\mathcal I}^{[n]}(\xi,t,\nuhat)
=-\int_0^1dy\,\frac{2}{(1+\xi y^2)^2}
\,f^{[n]}(y,t,\nuhat),
\label{eq:Itilde}
\end{equation}
where $\nuhat=\nu/p$ and
\begin{equation}
f^{[n]}=
\begin{cases}
\displaystyle
\frac{\sqrt{s+y^2-\nuhat^2+t^2-1}}{\sqrt{2}\,s}\,y^n,
& n\ \text{odd},\\[8pt]
\displaystyle
i\,\frac{\sqrt{2}\,\nuhat t}{s\sqrt{s+y^2-\nuhat^2+t^2-1}}\,y^{n+1},
& n\ \text{even},
\end{cases}
\end{equation}
with
\begin{equation}
s=s(y,t,\nuhat)=
\sqrt{\left(y^2-\nuhat^2+t^2-1\right)^2
+4y^2\nuhat^2t^2}.
\end{equation}
The BGK factor appearing below is
\begin{equation}
\mathcal W(0,\nuhat)=1-\nuhat\,\arccot\nuhat.
\end{equation}

\begin{widetext}
The five static mass scales are then
\begin{subequations}
\label{eq:massscales}
\begin{align}
m_{\alpha}^2={}&-\frac{\xi t\,\md^2}{2(1-t^2)}
\Big[t f_1(\xi)-(1-t^2+\nuhat^2)\widetilde{\mathcal I}^{[1]}
-2i\nuhat t\widetilde{\mathcal I}^{[2]}
+\widetilde{\mathcal I}^{[3]}\Big],\\
m_{\beta}^2={}&\frac{\md^2}{2\mathcal W(0,\nuhat)}
\Big[f_0(\xi)-i\nuhat\widetilde{\mathcal I}^{[0]}
-\xi t\widetilde{\mathcal I}^{[1]}\Big],\\
m_{\gamma}^2={}&\frac{\xi\md^2}{2(1-t^2)}
\Big[(1+t^2)f_1(\xi)-t(1-t^2+2\nuhat^2)\widetilde{\mathcal I}^{[1]}
-i\nuhat(1+3t^2)\widetilde{\mathcal I}^{[2]}
+2t\widetilde{\mathcal I}^{[3]}\Big],\\
m_{\delta}^2={}&\frac{\md^2}{2\sqrt{1-t^2}}
\Big[\nuhat t f_0(\xi)-i\nuhat^2 t\widetilde{\mathcal I}^{[0]}
+\nuhat(1-\xi t^2)\widetilde{\mathcal I}^{[1]}
-i\xi t\widetilde{\mathcal I}^{[2]}\Big],\\
m_{\rho}^2={}&-\frac{\md^2}{2\mathcal W(0,\nuhat)\sqrt{1-t^2}}
\Big[\xi\nuhat\widetilde{\mathcal I}^{[1]}
+i\xi t\widetilde{\mathcal I}^{[2]}\Big].
\end{align}
\end{subequations}
Here every $\widetilde{\mathcal I}^{[n]}$ carries the common arguments $(\xi,t,\nuhat)$.  Although individual terms in Eq.~\eqref{eq:massscales} are written in complex form, the static combinations defining the mass scales are real.
\end{widetext}

The two effective screening masses entering the longitudinal response are most compactly written as
\begin{align}
\mathcal S&=m_\alpha^2+m_\beta^2+m_\gamma^2,\nonumber\\
\mathcal D&=m_\beta^2(m_\alpha^2+m_\gamma^2)-m_\delta^2m_\rho^2,\nonumber\\
m_\pm^2&=\frac12\left[\mathcal S\pm\sqrt{\mathcal S^2-4\mathcal D}\right].
\label{eq:mplusminus}
\end{align}

For the fluctuation function we use the unnormalized anisotropic angular average
\begin{equation}
\langle X\rangle_\xi
=\int\frac{d\Omega_v}{4\pi}\,
\frac{X(\bm v)}{\left[1+\xi(\bm v\!\cdot\!\bm n)^2\right]^{3/2}}.
\label{eq:avgxi}
\end{equation}
Let $p_\perp=p\sqrt{1-t^2}$ and
$\widetilde{\bm n}=\bm n-t\hat{\bm p}$, so that
$\widetilde n^{\,2}=1-t^2$.  For the finite collision rate used throughout this work, the static fluctuation function obtained from the Feynman propagator in Ref.~\cite{Debnath2026} is
\begin{widetext}
\begin{equation}
\begin{aligned}
\Phi(p,\xi,\nu;t)={}&
-\frac{2\lambda\md^2p}
{\left[(p^2+m_\beta^2)(p^2+m_\alpha^2+m_\gamma^2)-m_\delta^2m_\rho^2\right]^2}
\\[5pt]
&\times\left[
(p^2+m_\alpha^2+m_\gamma^2)^2
\frac{\langle1\rangle_\xi
\left\langle\dfrac{\nuhat}{(\bm v\cdot\hat{\bm p})^2+\nuhat^2}\right\rangle_\xi}
{\left\langle\dfrac{(\bm v\cdot\hat{\bm p})^2}{(\bm v\cdot\hat{\bm p})^2+\nuhat^2}\right\rangle_\xi}
+\frac{p^2m_\delta^2m_\rho^2}{p_\perp^2}
\left\langle
\frac{\nuhat(\bm v\cdot\widetilde{\bm n})^2}
{(\bm v\cdot\hat{\bm p})^2+\nuhat^2}
\right\rangle_\xi
\right].
\end{aligned}
\label{eq:PhiExplicit}
\end{equation}
\end{widetext}
where the hard scale $\lambda$ is identified with $T$ in the present calculation, as stated above.  In the isotropic limit, $m_\alpha=m_\gamma=m_\delta=m_\rho=0$ and $m_\beta=\md$.  The dependence on the propagation angle $t$ then drops out, and Eq.~\eqref{eq:PhiExplicit} reduces to
\begin{align}
\left.\Phi(p,\xi,\nu;t)\right|_{\xi=0}
&\equiv \Phi_{\rm iso}(p,\nu)\nonumber\\
&=-\frac{2Tp\md^2}{(p^2+\md^2)^2}
\frac{\arccot\nuhat}{1-\nuhat\arccot\nuhat}.
\label{eq:PhiIso}
\end{align}
which provides a useful check of the anisotropic implementation.

Defining
\begin{equation}
{\cal R}(p,t)=\frac{p^2+m_\alpha^2+m_\gamma^2}
{(p^2+m_+^2)(p^2+m_-^2)},
\end{equation}
and writing $x=pr$, the angular kernel entering both the real and
imaginary parts of the potential is
\begin{equation}
\mathcal K(x;t,\mu)=\cos(xt\mu)
J_0\!\left(x\sqrt{1-t^2}\sqrt{1-\mu^2}\right).
\label{eq:kernel}
\end{equation}
The static response is even under $t\to -t$, so the contributions from
$t$ and $-t$ can be combined and the angular integration may be written
over $0\le t\le1$.  This symmetry is the origin of the cosine factor in
Eq.~\eqref{eq:kernel}.

The real part of the potential is evaluated with the ultraviolet subtraction
introduced in the potential construction~\cite{Debnath2026},
\begin{align}
V_R(r,\mu)=&
-\lim_{\Lambda_{\rm UV}\to\infty}\frac{2}{\pi}
\int_0^{\Lambda_{\rm UV}}dp\int_0^1dt\,p^2
[\mathcal K(pr;t,\mu)-1]\nonumber\\
&\times\left(\alpha+\frac{2\sigma}{p^2}\right){\cal R}(p,t)
+V_{\rm UV,sub}(\Lambda_{\rm UV}),
\label{eq:VR}
\end{align}
where the subtraction term is
\begin{equation}
V_{\rm UV,sub}(\Lambda_{\rm UV})
=-\frac{2\alpha}{\pi}\int_0^{\Lambda_{\rm UV}}dp
=-\frac{2\alpha}{\pi}\Lambda_{\rm UV}.
\label{eq:VUVsub}
\end{equation}
It removes the linear ultraviolet divergence of the Coulomb contribution;
the finite result is independent of $\Lambda_{\rm UV}$ once the cutoff is
taken sufficiently large.

The imaginary part is
\begin{align}
V_I(r,\mu)=&-\frac{2}{\pi}\int_0^\infty dp\int_0^1dt\,
[\mathcal K(pr;t,\mu)-1]\nonumber\\
&\times\left(\alpha+\frac{2\sigma}{p^2}\right)\Phi(p,\xi,\nu;t),
\label{eq:VI}
\end{align}
with the static fluctuation function $\Phi$ given explicitly in Eq.~\eqref{eq:PhiExplicit}.

\subsection{Pinch-free domain}
\label{sec:pinch}

The imaginary potential contains a factor $(p^2+m_-^2)^{-2}$.  A real pole occurs when $m_-^2$ is real and negative and the condition $p^2=-m_-^2$ is satisfied on the integration contour.  Since the static mass scales depend on $p$ and $\nu$ through $\nuhat=\nu/p$, the boundary can be written as
\begin{equation}
\left(\frac{\nu_{\rm crit}(\xi)}{\md}\right)^2
=\max_{t,\nuhat}
\left[-\nuhat^2\frac{m_-^2(\xi,t,\nuhat)}{\md^2}\right]_{m_-^2<0,\,m_-^2\in\mathbb R}.
\label{eq:nucrit}
\end{equation}
The resulting critical ratio is independent of temperature by dimensional analysis and agrees with the singular region identified in Ref.~\cite{Debnath2026}.  Inverting Eq.~\eqref{eq:nu} at this boundary gives
\begin{equation}
c_{\min}(T,\xi)=
\exp\!\left[\frac{\nu_{\rm crit}(\xi)/\md}{C_\nu\,\alpha_s(T)^{3/2}}\right]
-\frac{0.25}{\alpha_s(T)}.
\label{eq:cmin}
\end{equation}
The representative boundary values used in the calculation are listed in Table~\ref{tab:nucrit}.  Together, Eqs.~\eqref{eq:cmin} and \eqref{eq:cscheduledef} define the collision-rate band used at each anisotropy.

\begin{table}[hbt]
\caption{Minimum $\nu/\md$ required for a pinch-free imaginary potential.}
\label{tab:nucrit}
\centering
\begin{ruledtabular}
\begin{tabular}{c|cccccc}
$\xi$ & 0.1 & 0.5 & 1 & 2 & 5 & 10\\
\hline
$\nu_{\rm crit}/\md$ & 0.078 & 0.152 & 0.188 & 0.217 & 0.236 & 0.234\\
\end{tabular}
\end{ruledtabular}
\end{table}

\section{Complex Schr\"odinger equation}
\label{sec:sch}

We subtract only the real continuum threshold
$\VinfR\equiv\lim_{r\to\infty}V_R(r,\mu)$ and solve
\begin{equation}
\left[-\frac{\nabla^2}{2\mu_Q}+V_R(\bm r)+iV_I(\bm r)-\VinfR\right]\Psi_n(\bm r)
=\Ecal_n\Psi_n(\bm r),
\label{eq:schrodinger}
\end{equation}
where the asymptotic threshold is independent of $\mu$ because all
nonzero angular multipoles vanish for $r\to\infty$.  Here $\mu_Q=m_Q/2$; we take
$m_c=1.275$ GeV and $m_b=4.18$ GeV.  The complex eigenvalue is written as
\begin{equation}
\Ecal_n=E_{R,n}-\frac{i}{2}\Gamma_n,
\end{equation}
and the binding energy and thermal width are defined by
\begin{equation}
E_{B,n}=-E_{R,n},\qquad \Gamma_n=-2\,\mathrm{Im}\,\Ecal_n.
\label{eq:EBGamma}
\end{equation}
Thus $E_{R,n}=\mathrm{Re}\,\Ecal_n$ is the real energy of state $n$
measured relative to the continuum threshold.  For a bound state
$E_{R,n}<0$, so $E_{B,n}>0$ is the corresponding binding energy, while
$\Gamma_n$ is its thermal decay width.  The main operational dissociation criterion is
\begin{equation}
\Gamma_n(T_d)=2E_{B,n}(T_d),
\label{eq:Tdcrit}
\end{equation}
while $E_B=0$ and $\Gamma=E_B$ are retained as auxiliary criteria.

\subsection{Axisymmetric two-dimensional problem and coupled partial waves}
\label{sec:coupledPW}

At finite anisotropy the potential depends on both the separation $r$ and
the angle $\Theta$ between the quark--antiquark axis and the anisotropy
direction. Thus the Schr\"odinger equation is an axisymmetric
two-dimensional problem in $(r,\Theta)$; it is not reduced by replacing
the potential with an angular average. We solve the angular dependence
spectrally. With $\mu=\cos\Theta$, the potential is expanded in terms of
Legendre multipoles as
\begin{align}
V(r,\mu)&=\sum_{L=0,2,4,\ldots}V_L(r)P_L(\mu),\nonumber\\
V_L(r)&=\frac{2L+1}{2}\int_{-1}^{1}d\mu\,P_L(\mu)V(r,\mu).
\label{eq:Vmult}
\end{align}
Only even $L$ occur because the spheroidal medium is invariant under
$\hat{\bm n}\to-\hat{\bm n}$, or equivalently $\mu\to-\mu$.

Axial symmetry also implies that the magnetic quantum number $m$ is
conserved. For a fixed $m$ we therefore expand the right eigenfunction as
\begin{equation}
\Psi_{nm}(r,\Theta,\phi)
=\frac{1}{r}\sum_{\ell\geq |m|}
u_{n\ell m}(r)Y_{\ell m}(\Theta,\phi),
\label{eq:psiPW}
\end{equation}
where
\begin{equation}
Y_{\ell m}(\Theta,\phi)
={\cal N}_{\ell m}P_\ell^m(\cos\Theta)e^{im\phi}.
\label{eq:YlmTheta}
\end{equation}
Here $P_\ell^m$ is the associated Legendre function and
\[
{\cal N}_{\ell m}
=\left[\frac{2\ell+1}{4\pi}
\frac{(\ell-m)!}{(\ell+m)!}\right]^{1/2}
\]
is the spherical-harmonic normalization, with the conventional
Condon--Shortley phase included in $P_\ell^m$. The angular structure of the
eigenfunction is represented by the associated Legendre functions in the
spherical harmonics, while the radial amplitudes $u_{n\ell m}(r)$ determine
the weight of each partial-wave component.

Substituting Eqs.~\eqref{eq:Vmult} and \eqref{eq:psiPW} into the full
Schr\"odinger equation, multiplying from the left by
$Y_{\ell m}^*(\Theta,\phi)$, and integrating over $d\Omega$ gives the
coupled radial equations
\begin{align}
&\left[
-\frac{1}{2\mu_Q}\frac{d^2}{dr^2}
+\frac{\ell(\ell+1)}{2\mu_Qr^2}
\right]u_{n\ell m}(r)
\nonumber\\
&\quad+
\sum_{\ell'}\sum_{L}
\widetilde V_L(r)\,
C_{\ell\ell'}^{Lm}\,
u_{n\ell' m}(r)
=
{\cal E}_{nm}\,u_{n\ell m}(r),
\label{eq:coupledRadial}
\end{align}
where $\widetilde V_0(r)=V_0(r)-V_\infty^R$ and
$\widetilde V_{L>0}(r)=V_L(r)$. The angular coupling matrix is
\begin{equation}
C_{\ell\ell'}^{Lm}
=\int d\Omega\,
Y_{\ell m}^*(\Omega)P_L(\cos\Theta)Y_{\ell'm}(\Omega).
\label{eq:couplings}
\end{equation}
Equivalently,
\begin{align}
C_{\ell\ell'}^{Lm}
={}&(-1)^m\sqrt{(2\ell+1)(2\ell'+1)}
\begin{pmatrix}
\ell & L & \ell'\\
0 & 0 & 0
\end{pmatrix}
\nonumber\\
&\times
\begin{pmatrix}
\ell & L & \ell'\\
-m & 0 & m
\end{pmatrix}.
\label{eq:couplings3j}
\end{align}
The coefficients $C_{\ell\ell'}^{Lm}$ are the matrix elements of the
Legendre multipoles in the spherical-harmonic basis. Consequently, each
$V_L(r)$ contributes to the diagonal and, when allowed by the angular
selection rules, off-diagonal blocks of the coupled radial Hamiltonian.
The Wigner-$3j$ representation gives the selection rules: $m$ is conserved,
$|\ell-\ell'|\leq L\leq \ell+\ell'$, and
$\ell+\ell'+L$ must be even. Since only even $L$ occur, even-$\ell$
and odd-$\ell$ sectors do not mix with one another.

For example, the quadrupole component $V_2$ couples the $S$ and $D$
channels through
\begin{equation}
C_{02}^{\,2\,0}=\frac{1}{\sqrt{5}},
\end{equation}
and hence generates the $S$--$D$--$G$ admixture of an $S$-like state.
After the angular projection, the eigenvalue problem is expressed in terms
of the coupled radial amplitudes $u_{n\ell m}(r)$. The angular information
is retained through the spherical-harmonic basis and the matrices
$C_{\ell\ell'}^{Lm}$, so that the procedure is a spectral representation
of the original $(r,\Theta)$ problem rather than an angular average. The
angular dependence of the eigenstate can be reconstructed at any $r$ from
\begin{equation}
r^2|\Psi_{nm}(r,\Theta,\phi)|^2
=
\left|
\sum_\ell
u_{n\ell m}(r)\,
{\cal N}_{\ell m}P_\ell^m(\cos\Theta)
\right|^2,
\label{eq:angularDensity}
\end{equation}
where the common phase $e^{im\phi}$ drops out of the probability density.
The angle-integrated radial density used below follows by integrating this
quantity over the angular variables.

At $\xi=0$ all $V_{L>0}$ vanish and
$C_{\ell\ell'}^{0m}=\delta_{\ell\ell'}$, so
Eq.~\eqref{eq:coupledRadial} reduces to independent one-dimensional
radial equations and the spectrum becomes independent of $m$. At
finite $\xi$ the coupled system retains the full axisymmetric angular
structure. In practice we truncate the potential multipoles and the
partial-wave basis only after checking convergence, as discussed below.

\section{Numerical implementation}
\label{sec:numerics}

For each $(T,\xi,c)$ point, the Legendre multipoles of the complex
anisotropic potential are first evaluated and combined with the angular
matrices $C_{\ell\ell'}^{Lm}$ in Eq.~\eqref{eq:coupledRadial}. The resulting
sparse block Hamiltonian is the spherical-harmonic spectral representation of
the axisymmetric two-dimensional Schr\"odinger operator: every diagonal block
contains the radial kinetic and centrifugal terms, while the multipoles
$V_{L>0}$ generate the off-diagonal partial-wave couplings. The full
non-Hermitian block eigenvalue problem is then solved for each heavy flavor
and fixed-$m$ sector. Along continuous temperature trajectories, adjacent
eigenstates are associated by their wave-function overlap so that the
physical branch can be followed through the medium scan.
Independent medium and channel sectors are evaluated in parallel, while the
temperature ordering is preserved within each trajectory.  The production scan
extends down to $T=0.15$ GeV so that the calculated widths overlap the lowest-temperature
lattice points used below.  These lowest temperatures are close to the crossover region, where the perturbative medium input is least controlled, and are therefore used mainly for comparison with the lattice results.

For each state we extract the complex pole, binding energy, thermal width, radial observables, and partial-wave content.  A first-order diagnostic is evaluated at the same medium point so that the full pole width can be compared with
\begin{equation}
\Gamma_n^{\rm PT}=-2\langle\psi_n^{(R)}|V_I|\psi_n^{(R)}\rangle,
\label{eq:gammaPT}
\end{equation}
where $\psi_n^{(R)}$ is obtained from the real Hamiltonian at the same medium
point.

We checked the numerical implementation against several limiting cases.  At $\xi=0$, the numerical real potential agrees with the analytic screened-Cornell form, the continuum threshold matches the corresponding analytic result, and the higher Legendre multipoles vanish within numerical precision.  We also repeated representative calculations after independently refining the radial grid, the momentum and angular integrations, the Legendre expansion of the potential, and the partial-wave basis.  Across these tests the change in $T_d$ remained below $0.02\%$, while the changes in $E_B(T_d)$ and $\Gamma(T_d)$ were comparable or smaller.  These numerical variations are much smaller than the medium-dependent effects discussed below.

\section{Results and discussion}
\label{sec:results}

\subsection{Medium-modified radial probability density}

The complex eigenvalue problem also gives the in-medium right eigenfunction.
For visualizing its spatial extent, we normalize the right eigenfunction with
the standard $L^2$ norm and define the angle-integrated radial density as
\begin{equation}
P_{nm}(r)\equiv r^2\int d\Omega\,
\left|\Psi_{nm}(r,\Omega)\right|^2.
\label{eq:PrDefinition}
\end{equation}
Using the partial-wave expansion in Eq.~\eqref{eq:psiPW}, this becomes
\begin{align}
P_{nm}(r)
={}&\sum_{\ell,\ell'}
 u_{n\ell m}(r)u^*_{n\ell' m}(r)
 \int d\Omega\,Y_{\ell m}(\Omega)Y^*_{\ell' m}(\Omega)
 \nonumber\\
={}&\sum_{\ell}\left|u_{n\ell m}(r)\right|^2,
\label{eq:PrPartialWaves}
\end{align}
where the second line follows from the orthonormality relation
$\int d\Omega\,Y_{\ell m}Y^*_{\ell' m}=\delta_{\ell\ell'}$.
The normalization of the full eigenfunction therefore implies
\begin{equation}
\int_0^\infty dr\,P_{nm}(r)
=\int d^3r\,|\Psi_{nm}(\bm r)|^2=1.
\label{eq:PrNorm}
\end{equation}

\begin{fullwidthhere}
\begin{center}
\includegraphics[width=0.98\textwidth]{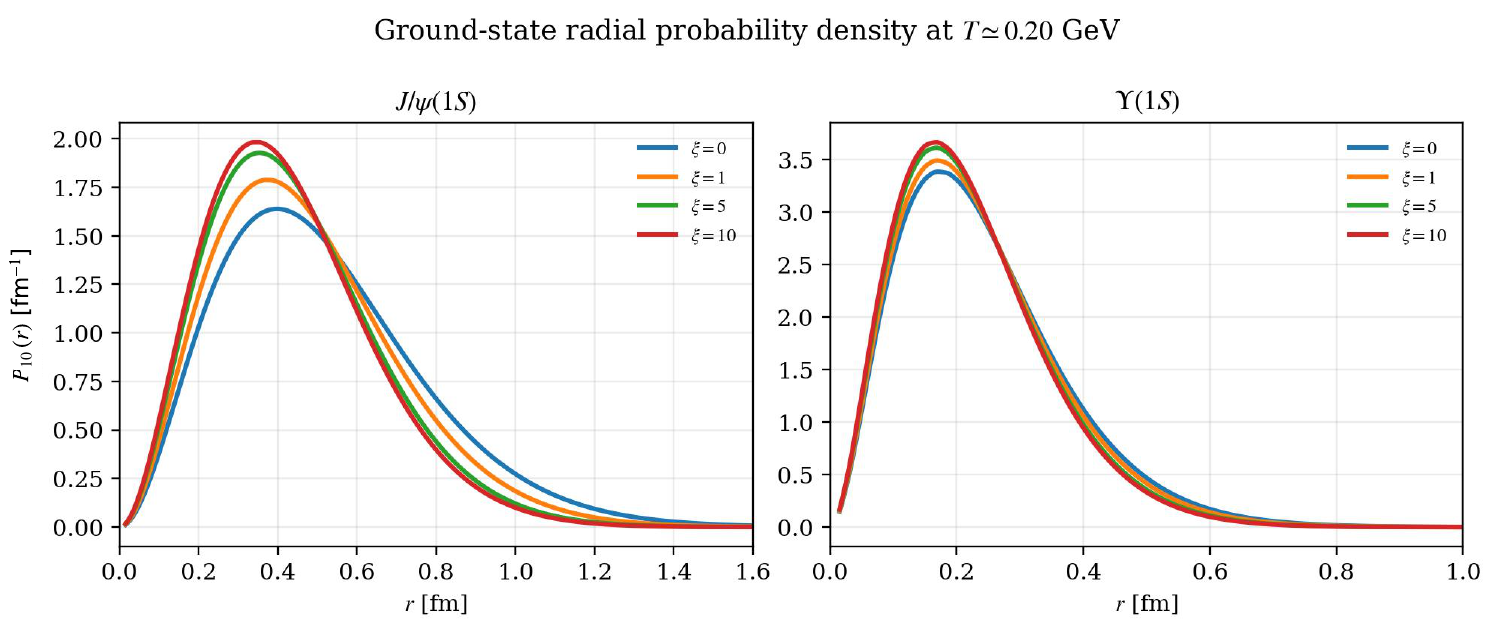}\par
\vspace{3pt}
\refstepcounter{figure}\label{fig:radialprob}
{\small\textbf{FIG.~\thefigure.} Normalized radial probability density for $J/\psi(1S)$ (left) and $\Upsilon(1S)$ (right) at $T\simeq0.20$ GeV.  Both panels show the $n=1$, $m=0$ sector; accordingly, the plotted quantity is denoted on the vertical axis by $P_{10}(r)$.  Each finite-$\xi$ curve uses the central pinch-safe collision parameter for that anisotropy.}
\end{center}
\end{fullwidthhere}

The radial density $P_{nm}(r)$ plotted in Fig.~\ref{fig:radialprob} is therefore
obtained only after the angular structure of the two-dimensional eigenfunction
has been integrated out.  For the $1S$ states displayed below, $n=1$ and
$m=0$, so the plotted quantity is $P_{1,0}(r)$.  The bottomonium
ground state is more compact, as expected from its larger reduced mass.
Increasing the anisotropy along the central pinch-safe trajectory shifts the
radial density toward smaller radii, with a stronger visible effect for
charmonium.  The change in the radial profile remains smooth, consistent with
the small higher-partial-wave admixtures discussed below.

\subsection{Isotropic collisional baseline}

\begin{figure*}[hbt]
\includegraphics[width=0.98\textwidth]{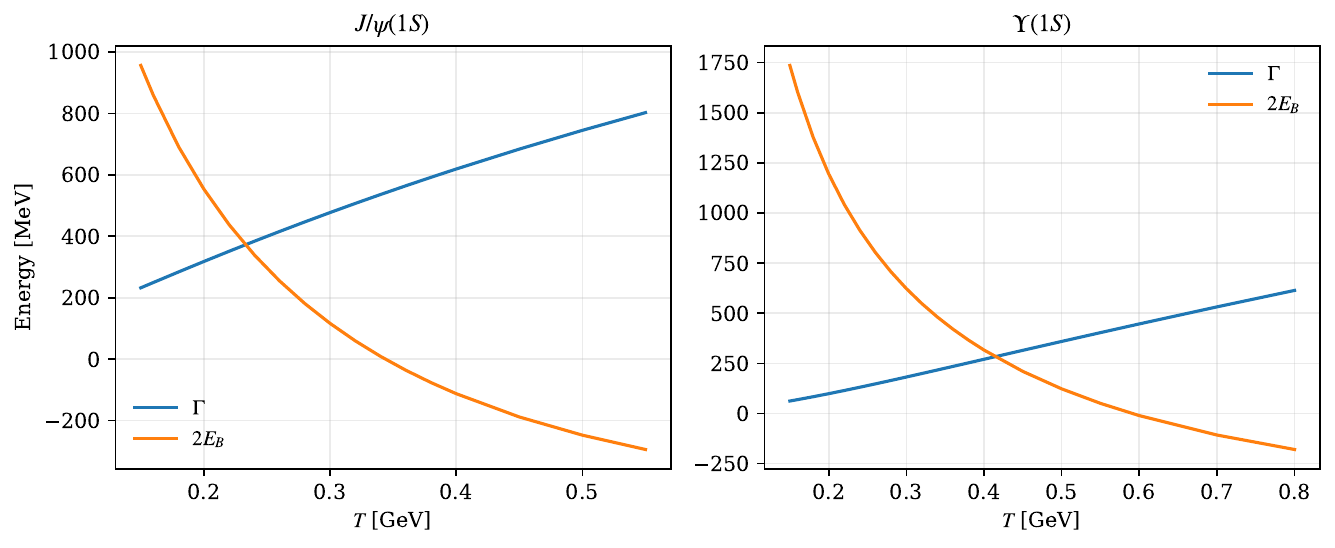}
\caption{Isotropic collisional baseline for the central value $c=1.5$.  The curves show the full complex-pole width $\Gamma$ and twice the binding energy $2E_B$ for $J/\psi(1S)$ (left) and $\Upsilon(1S)$ (right).  Their crossing defines the main dissociation temperature.}
\label{fig:isobase}
\end{figure*}

The $\xi=0$ calculation provides the reference point for the anisotropic analysis using the same running coupling, collision prescription, complex Hamiltonian, and dissociation criterion.  The crossings of $\Gamma(T)$ and $2E_B(T)$ for the $1S$ charmonium and bottomonium states are shown in Fig.~\ref{fig:isobase}.  Varying $c$ over the preferred isotropic interval produces only a modest shift of the crossing temperatures.

The isotropic limit also provides a useful check of the implementation: the calculated real potential reproduces the analytic screened-Cornell result, and the higher Legendre multipoles vanish within numerical precision.

\subsection{Comparison with lattice thermal widths}

We compare the calculated thermal widths with available lattice determinations.  Absolute quarkonium masses are not shown because this potential-model setup does not fix the additive heavy-quark rest-mass contribution without an additional matching prescription.

Figure~\ref{fig:hotqcd} compares the isotropic widths with the HotQCD results for the $1S$ charmonium and bottomonium channels~\cite{Ali:2025iux}.  For $\Upsilon(1S)$ we also include the lattice data of Larsen \textit{et al.}~\cite{Larsen:2019zqv}, labeled ``BNL'' in the legend.  The charmonium widths are broadly comparable to the lattice bands but do not reproduce their temperature dependence uniformly.  For bottomonium, the present widths tend to lie above the lattice determinations over much of the common temperature range.  The lattice results are used only as a comparison; they are not used to tune the collision parameter.

\begin{fullwidthhere}
\begin{center}
\includegraphics[width=0.98\textwidth]{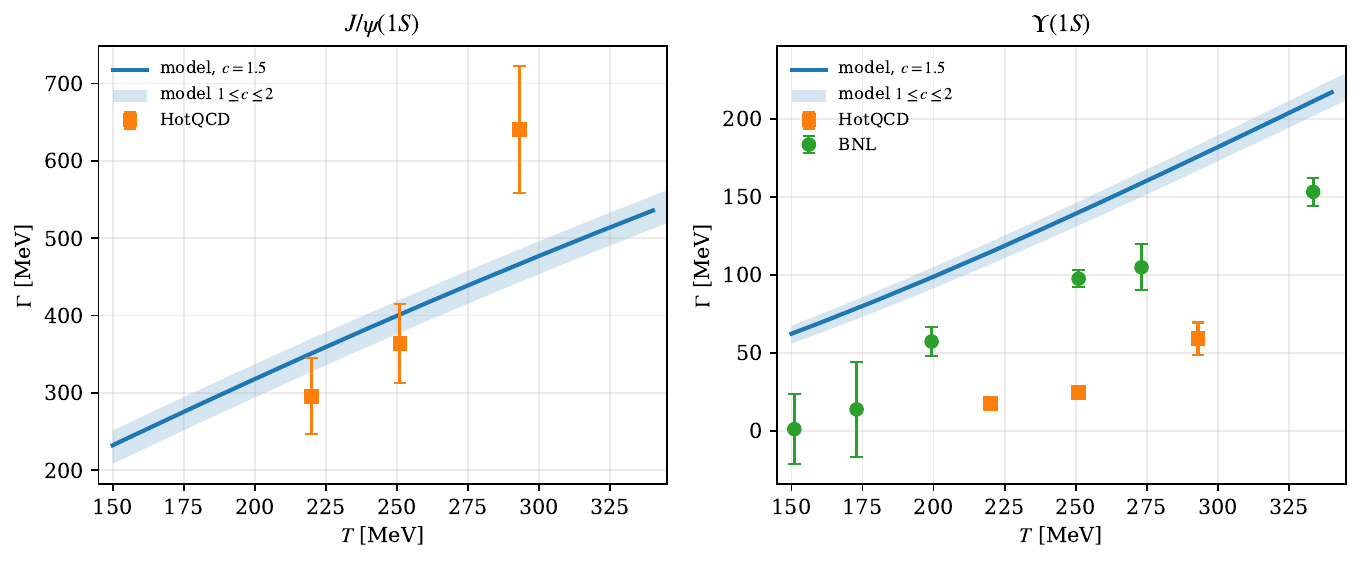}\par
\vspace{3pt}
\refstepcounter{figure}\label{fig:hotqcd}
{\small\textbf{FIG.~\thefigure.} Thermal widths in the isotropic medium compared with lattice results.  Left: $J/\psi(1S)$ with the HotQCD intervals~\cite{Ali:2025iux}.  Right: $\Upsilon(1S)$ with the HotQCD intervals~\cite{Ali:2025iux} and the Larsen \textit{et al.} data~\cite{Larsen:2019zqv}, labeled BNL.  The solid curve is the central $c=1.5$ result and the shaded region spans $1\leq c\leq2$.}
\end{center}
\end{fullwidthhere}

Figure~\ref{fig:hotqcdXi} places representative finite-$\xi$ trajectories on the same axes.  Since the lattice calculations refer to an equilibrium isotropic plasma, this is not a direct validation of the anisotropic result; it only shows the direction and size of the predicted shift relative to a common reference.  The charmonium curves move appreciably with $\xi$, whereas the bottomonium widths remain systematically high.

\begin{fullwidthhere}
\begin{center}
\includegraphics[width=0.98\textwidth]{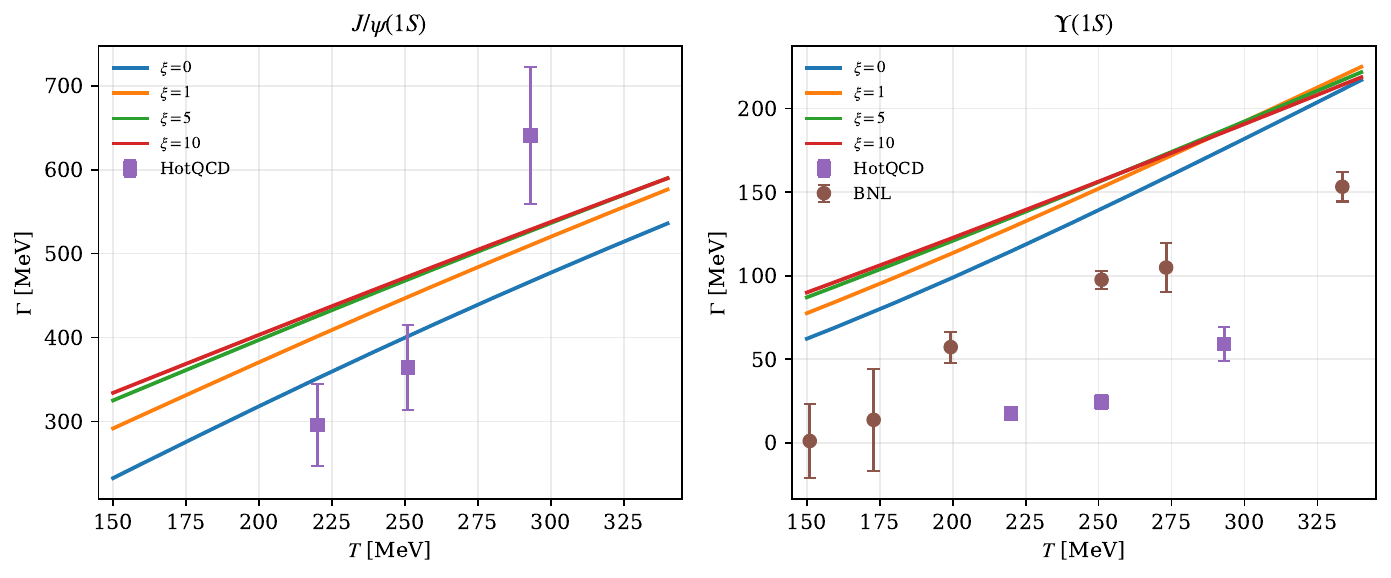}\par
\vspace{3pt}
\refstepcounter{figure}\label{fig:hotqcdXi}
{\small\textbf{FIG.~\thefigure.} Representative finite-anisotropy width trajectories compared with the same isotropic lattice reference data.  Each curve uses the central pinch-safe collision parameter for the corresponding $\xi$.  The $J/\psi(1S)$ panel contains the HotQCD intervals~\cite{Ali:2025iux}; the $\Upsilon(1S)$ panel contains the HotQCD intervals~\cite{Ali:2025iux} and the BNL data~\cite{Larsen:2019zqv}.  The finite-$\xi$ curves should be viewed as a diagnostic comparison to an isotropic lattice reference.}
\end{center}
\end{fullwidthhere}

\subsection{Collision-rate band used at finite anisotropy}

The collision parameters used in the finite-$\xi$ calculation are summarized in Table~\ref{tab:cschedule}.  The preferred interval $[1,1.5,2]$ is retained at weak anisotropy; at larger $\xi$ it is shifted upward only as required by the pinch-free condition.

\begin{table}[hbt]
\caption{Pinch-safe collision-parameter bands used in the finite-anisotropy calculation.}
\label{tab:cschedule}
\centering
\begin{ruledtabular}
\begin{tabular}{c|ccc}
$\xi$ & $c_{\rm low}$ & $c_{\rm cen}$ & $c_{\rm high}$\\
\hline
0    & 1.000 & 1.500 & 2.000\\
0.1  & 1.000 & 1.500 & 2.000\\
0.5  & 2.635 & 3.135 & 3.635\\
1    & 4.043 & 4.543 & 5.043\\
2    & 5.566 & 6.066 & 6.566\\
5    & 6.787 & 7.287 & 7.787\\
10   & 6.670 & 7.170 & 7.670\\
\end{tabular}
\end{ruledtabular}
\end{table}

\subsection{Finite-anisotropy $1S$ spectrum and dissociation}

The main results for the two ground states are collected in Table~\ref{tab:1Sresults}.  The uncertainty interval on $T_d$ is the envelope obtained from the lower and upper members of the corresponding pinch-safe $c$ band.  The central value always refers to the central member listed in Table~\ref{tab:cschedule}.

\begin{table*}[hbt]
\caption{Ground-state dissociation results from the full complex Hamiltonian using $\Gamma=2E_B$.  The asymmetric errors on $T_d$ represent the lower/upper $c$ trajectories, not a statistical uncertainty.  Energies in the last four columns are evaluated at the central crossing.}
\label{tab:1Sresults}
\centering
\begin{ruledtabular}
\begin{tabular}{c c c c c c c c}
$\xi$ & $c_{\rm cen}$ &
$T_d^{J/\psi}$ [GeV] & $E_B^{J/\psi}$ [GeV] & $\Gamma_{J/\psi}$ [GeV] &
$T_d^\Upsilon$ [GeV] & $E_B^\Upsilon$ [GeV] & $\Gamma_\Upsilon$ [GeV]\\
\hline
0   & 1.500 & $0.2332_{-0.0046}^{+0.0058}$ & 0.1864 & 0.3728 & $0.4152_{-0.0043}^{+0.0053}$ & 0.1420 & 0.2841\\
0.1 & 1.500 & $0.2373_{-0.0046}^{+0.0061}$ & 0.1868 & 0.3737 & $0.4224_{-0.0043}^{+0.0053}$ & 0.1423 & 0.2846\\
0.5 & 3.135 & $0.2382_{-0.0027}^{+0.0032}$ & 0.2057 & 0.4113 & $0.4337_{-0.0027}^{+0.0030}$ & 0.1509 & 0.3017\\
1   & 4.543 & $0.2442_{-0.0022}^{+0.0024}$ & 0.2190 & 0.4380 & $0.4483_{-0.0020}^{+0.0023}$ & 0.1574 & 0.3148\\
2   & 6.066 & $0.2569_{-0.0017}^{+0.0018}$ & 0.2346 & 0.4691 & $0.4756_{-0.0019}^{+0.0020}$ & 0.1666 & 0.3332\\
5   & 7.287 & $0.2866_{-0.0017}^{+0.0019}$ & 0.2590 & 0.5181 & $0.5310_{-0.0019}^{+0.0020}$ & 0.1825 & 0.3649\\
10  & 7.170 & $0.3186_{-0.0020}^{+0.0022}$ & 0.2810 & 0.5620 & $0.5881_{-0.0021}^{+0.0022}$ & 0.1982 & 0.3963\\
\end{tabular}
\end{ruledtabular}
\end{table*}

\begin{figure}[hbt]
\includegraphics[width=0.98\columnwidth]{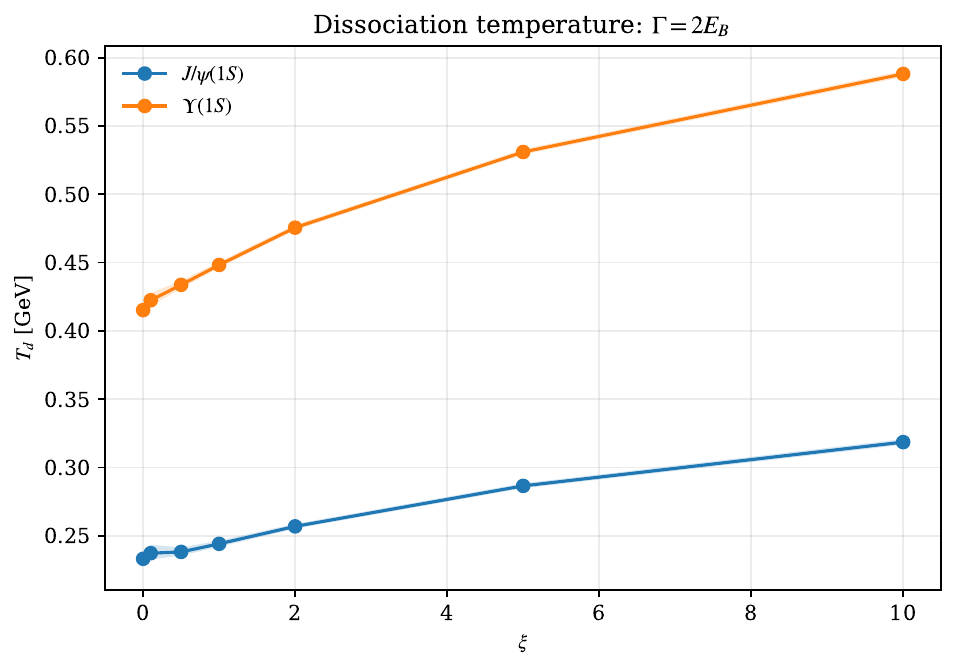}
\caption{Dissociation temperature versus momentum-space anisotropy for the $1S$ charmonium and bottomonium states.  The solid curves show the central pinch-safe trajectory and the narrow bands are generated from the lower and upper members of the continued $c$ interval.}
\label{fig:TdXi}
\end{figure}

Figure~\ref{fig:TdXi} shows the dissociation temperatures as a function of $\xi$.  For both ground states, $T_d$ increases with anisotropy along the central pinch-safe trajectory.  At larger $\xi$ this trend reflects both the direct anisotropy dependence of the potential and the collision rate required to remain in the pinch-free domain; it should therefore not be interpreted as a fixed-collision-rate anisotropy scan.

The binding energy at the crossing also increases with anisotropy, and the corresponding width follows from the condition $\Gamma(T_d)=2E_B(T_d)$.  The numerical values are collected in Table~\ref{tab:1Sresults}.

\subsection{Full complex pole versus first-order treatment}

The first-order diagnostic uses an eigenstate of the real Hamiltonian at the same medium point and therefore isolates the effect of treating $V_I$ nonperturbatively, without introducing the additional Coulomb-wave-function approximation used in the earlier width estimate.  This construction is closely related to Ref.~\cite{Srivastava:2018vxp}, where temperature-dependent wave functions were obtained from the real potential and the width was then evaluated as a matrix element of $V_I$.  The comparison in Fig.~\ref{fig:fullPT} shows that the full complex width is smaller than the first-order result over the scanned range.

\begin{fullwidthhere}
\begin{center}
\includegraphics[width=0.98\textwidth]{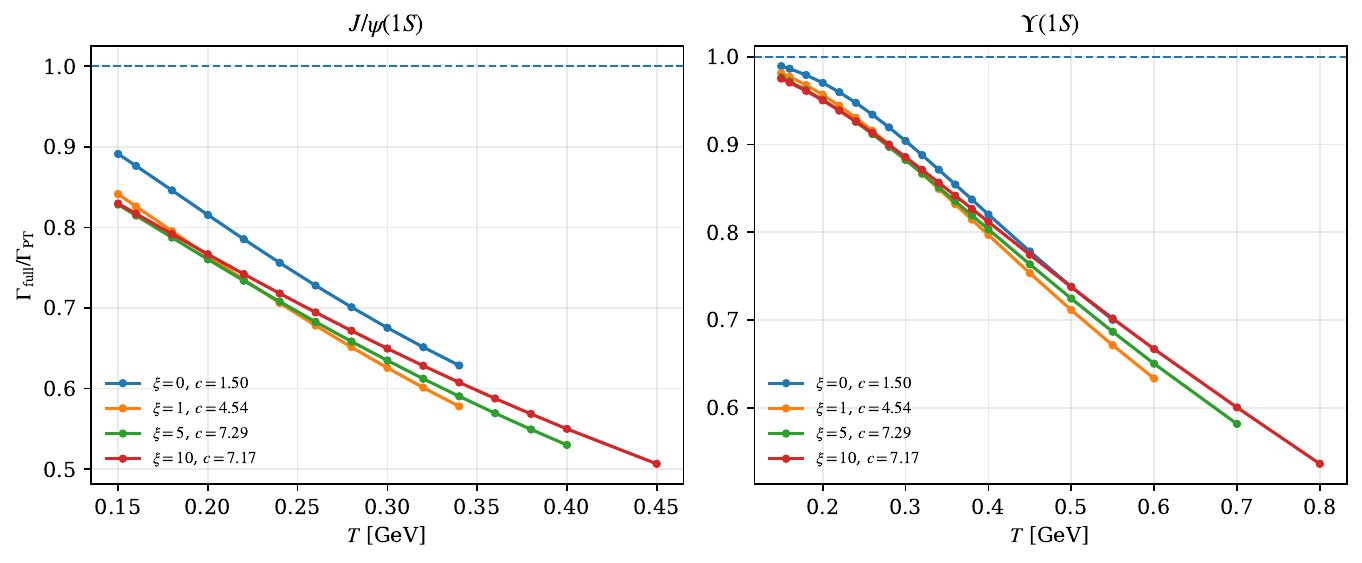}\par
\vspace{3pt}
\refstepcounter{figure}\label{fig:fullPT}
{\small\textbf{FIG.~\thefigure.} Ratio ${\cal R}_\Gamma=\Gamma_{\rm full}/\Gamma_{\rm PT}$ for the ground-state charmonium (left) and bottomonium (right) channels.  Each curve uses the central pinch-safe collision parameter for that anisotropy; the actual $c$ value is shown in the legend.  Only well-matched bound states are plotted.}
\end{center}
\end{fullwidthhere}

Near dissociation the difference is at the tens-of-percent level, especially for charmonium.  At fixed temperature, however, ${\cal R}_\Gamma$ is not strictly monotonic in $\xi$: over part of the common range the $\xi=10$ curve lies above the $\xi=5$ curve.  The two trajectories have very similar collision parameters because the pinch boundary has nearly saturated by $\xi=5$--10, so this ordering is not driven primarily by a large change in $c$.  Instead, the stronger binding at $\xi=10$ changes the real-state wave function and the relative size of the nonlinear imaginary-potential correction; $\Gamma_{\rm full}$ and $\Gamma_{\rm PT}$ therefore do not scale identically with anisotropy.  At the respective dissociation points, which occur at different temperatures, the ratio still decreases as $\xi$ increases.  Thus the nonmonotonic fixed-temperature ordering is a feature of the ratio rather than a state-matching artifact.

\subsection{Angular admixture of the $1S$ states}

The partial-wave decomposition provides a direct measure of the angular
structure induced by the noncentral interaction.  For the $1S$ sector,
$m=0$ and only even partial waves occur, so the eigenfunction has the form
\begin{align}
\Psi_{10}(r,\Theta)
=\frac{1}{r}\Big[&u_{100}(r)Y_{00}(\Theta)
+u_{120}(r)Y_{20}(\Theta)\nonumber\\
&+u_{140}(r)Y_{40}(\Theta)+\cdots\Big].
\label{eq:1Sadmixture}
\end{align}
For an $L^2$-normalized right eigenfunction, the weight of a given partial
wave is
\begin{equation}
w_\ell=\int_0^\infty dr\,|u_{1\ell0}(r)|^2,
\qquad \sum_{\ell=0,2,4,\ldots}w_\ell=1.
\label{eq:partialWeights}
\end{equation}
At the central dissociation point the state remains overwhelmingly
$S$-wave dominated throughout the anisotropy range considered here.  Even
at $\xi=10$, the $D$-wave weight remains below $8.3\times10^{-4}$, while
the $G$-wave weight is of order $10^{-6}$ or smaller.  The anisotropic
interaction therefore changes the complex pole and the radial profile of
the $1S$ states more visibly than it changes their partial-wave
composition.

The present phenomenological analysis is restricted to the $1S$
charmonium and bottomonium states.  A quantitative extension to physical
$P$-wave quarkonia would require the spin-dependent interactions responsible
for the vacuum fine structure, together with their modification in the
anisotropic medium; these terms are not included in the Hamiltonian used
here.

\section{From static pole widths to survival factors and suppression observables}

The complex pole gives an in-medium decay rate, but not by itself a nuclear modification factor.  As a simple illustration, we define a static-medium survival probability.  For a state
held at fixed $(T,\xi,c)$ for a time interval $\Delta\tau$, the probability
of remaining in that state is
\begin{equation}
S_n(T,\xi,c;\Delta\tau)=
\exp\!\left[-\frac{\Gamma_n(T,\xi,c)\,\Delta\tau}{\hbar c}\right],
\label{eq:Sstatic}
\end{equation}
when $\Gamma$ is expressed in GeV and $\Delta\tau$ in fm/$c$.  This quantity is not an $R_{AA}$ prediction; it only illustrates the effect of the calculated width for a fixed exposure time.

Table~\ref{tab:survival} gives this benchmark for the two $1S$ states at
$T=0.20$ GeV and $\Delta\tau=1$ fm/$c$.  For each $\xi$, the central value
uses $c_{\rm cen}$ and the bracket gives the envelope from $c_{\rm low}$ and
$c_{\rm high}$.  Figure~\ref{fig:survival}(a) displays the same information as
a function of anisotropy.  Figure~\ref{fig:survival}(b) shows the corresponding
exposure-time dependence at the same fixed temperature for the representative
choices $\xi=0$ and $10$, using the central pinch-safe trajectory.  The latter
is a static-medium time dependence: $T$, $\xi$, and $c$ are held fixed while
$\Delta\tau$ is varied.

\begin{table}[hbt]
\caption{Static-medium survival benchmark from Eq.~\eqref{eq:Sstatic} at $T=0.20$ GeV and $\Delta\tau=1$ fm/$c$.  The central value uses $c_{\rm cen}$ from Table~\ref{tab:cschedule}; brackets give the envelope from $c_{\rm low}$ and $c_{\rm high}$.}
\label{tab:survival}
\centering
\begin{ruledtabular}
\begin{tabular}{c cc}
$\xi$ & $S_{J/\psi}$ & $S_{\Upsilon}$\\
\hline
0   & 0.200 [0.181,0.225] & 0.606 [0.587,0.629]\\
0.1 & 0.205 [0.186,0.231] & 0.612 [0.594,0.636]\\
0.5 & 0.169 [0.159,0.180] & 0.578 [0.568,0.591]\\
1   & 0.153 [0.147,0.160] & 0.562 [0.554,0.571]\\
2   & 0.142 [0.138,0.147] & 0.551 [0.545,0.558]\\
5   & 0.134 [0.130,0.138] & 0.542 [0.536,0.548]\\
10  & 0.130 [0.126,0.134] & 0.537 [0.531,0.544]\\
\end{tabular}
\end{ruledtabular}
\end{table}

At this reference temperature the bottomonium ground state has a larger survival probability than charmonium because of its smaller thermal width.  At stronger anisotropy the survival factor decreases along the pinch-safe trajectories.  As elsewhere in the finite-$\xi$ analysis, this combines the anisotropy dependence of the potential with the accompanying collision-rate prescription.

\begin{fullwidthhere}
\begin{center}
\includegraphics[width=0.96\textwidth]{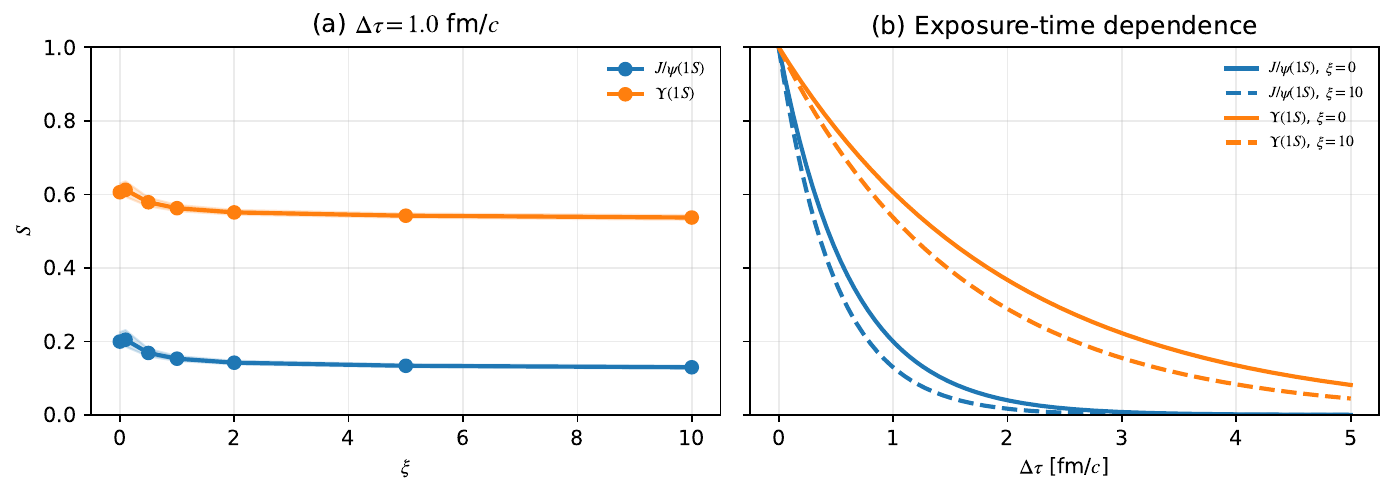}\par
\end{center}
\refstepcounter{figure}\label{fig:survival}
\begingroup\small
\noindent\textbf{FIG.~\thefigure.}~Static-medium survival probabilities for $J/\psi(1S)$ and $\Upsilon(1S)$ at $T=0.20$ GeV. (a) Anisotropy dependence at $\Delta\tau=1$ fm/$c$; bands indicate the $c_{\rm low}$--$c_{\rm high}$ range. (b) Exposure-time dependence for $\xi=0$ and $10$ along the central trajectories.
\par\endgroup
\end{fullwidthhere}

For a specified dynamical background, Eq.~\eqref{eq:Sstatic} generalizes to
the adiabatic survival factor
\begin{equation}
S_n=\exp\left[-\frac{1}{\hbar c}\int d\tau\,
\Gamma_n(T(\tau),\xi(\tau),c(\tau))\right].
\label{eq:Sdynamic}
\end{equation}
A phenomenological $R_{AA}$ additionally requires the production geometry,
feed-down, formation-time and cold-nuclear-matter effects, and regeneration
where relevant.  A calculation of $R_{AA}$ would therefore require a separate dynamical treatment.

\section{Summary and outlook}

We have solved the Schr\"odinger equation with the full complex heavy-quark potential of an anisotropic collisional QGP.  The calculation treats the noncentral interaction in a coupled partial-wave basis, so the complex pole and the medium-modified wave function are obtained consistently from the same Hamiltonian.

For the $1S$ charmonium and bottomonium states, the width-defined dissociation temperature rises with anisotropy along the pinch-free collision trajectories considered here.  The full complex-pole widths differ noticeably from first-order estimates based on $V_I$, with a larger effect for charmonium.  The $1S$ wave functions remain strongly dominated by the leading $S$-wave component, with only small $D$- and $G$-wave admixtures over the anisotropy range studied.

The comparison with lattice-QCD widths is mixed.  The charmonium result is of a similar overall scale to the HotQCD determination but does not reproduce its temperature dependence over the full common range.  The calculated bottomonium widths are generally larger than the HotQCD and BNL results.  These comparisons are not used to adjust either $c$ or $\xi$.

The static survival probability included here is only an illustration of how the calculated width accumulates over a finite time.  A phenomenological $R_{AA}$ requires a time-dependent medium, production geometry, feed-down, formation-time and cold-nuclear-matter effects, and, where relevant, regeneration.  Extending the present spectrum calculation to such an evolving background is left for future work.  A corresponding treatment of physical $P$-wave quarkonia would also require the spin-dependent interactions that generate the vacuum fine structure, together with their in-medium modification.

\begin{acknowledgments}
N.~H. acknowledges support from the Department of Atomic Energy, Government of India.
\end{acknowledgments}

\end{document}